# Limiter Control of a Chaotic RF Transistor Oscillator

Ned J. Corron, Buckley A. Hopper, and Shawn D. Pethel

U. S. Army Aviation and Missile Command

Redstone Arsenal, AL 35898

Abstract: We report experimental control of chaos in an electronic circuit at 43.9 MHz, which is the fastest chaos control reported in the literature to date. Limiter control is used to stabilize a periodic orbit in a tuned collector transistor oscillator modified to exhibit simply folded band chaos. The limiter is implemented using a transistor to enable monitoring the relative magnitude of the control perturbation. A plot of the relative control magnitude vs. limiter level shows a local minimum at period-1 control, thereby providing strong evidence that the controlled state is an unstable periodic orbit (UPO) of the uncontrolled system.

## 1. Introduction

Following the discovery of chaos control [Ott *et al.*, 1990], various researchers have speculated on potential engineering applications of chaos. A particular application that has been proposed is the use of chaotic waveforms for efficient data communications [Hayes *et al.*, 1993]. In this application, chaos control is used to actively encode a message signal in the symbolic dynamics of a chaotic oscillator. This concept was experimentally demonstrated using a chaotic audio-frequency circuit [Hayes *et al.*, 1994]. However, further development of this idea tacitly assumes that chaos control can be scaled to operate at the higher radio frequencies required for practical applications.

Unfortunately, the majority of approaches for closed-loop chaos control are not easily scalable to high-frequency systems. Two exceptions are delay feedback [Pyragas, 1992] and pulsewidth self-modulation techniques [Myneni *et al.*, 1999]. Indeed, Socolar *et al.* demonstrated stabilization of a driven 10.1 MHz chaotic circuit using extended time-delay autosynchronization [Socolar *et al.*, 1994], while Myneni *et al.* controlled a 19-MHz autonomous oscillator using pulsewidth modulation driven by the system transit time through a specified control window [Myneni *et al.*, 1999]. These experiments are the fastest closed-loop chaos controllers in the published literature to date. In both references, latency in the control feedback loop was identified as a critical limiting parameter. Recently, chaos control was demonstrated using simple limiters that provide effective closed loop control with truly minimal latency [Corron *et al.*, 2000]. In this letter, we report new experimental results using a diode limiter to control a chaotic RF transistor oscillator and stabilize a 43.9-MHz unstable periodic orbit (UPO).

The objective of chaos control as introduced by Ott *et al.* [Ott *et al.*, 1990] is to stabilize various unstable natural states of the uncontrolled chaotic oscillator. These waveforms include UPOs as well as aperiodic waveforms. Stabilizing UPOs may be beneficial in applications where chaotic behavior is undesirable; for example, chaos control has been proposed for cardiac defibrillation [Garfinkel *et al.*, 1992]. Stabilizing arbitrary dynamics may be useful for encoding symbolic dynamics for communications [Hayes *et al.*, 1993]. In either case, chaos control promises to be an efficient process—that is, the control perturbations required to stabilize these natural states are very small [Ott *et al.*, 1990]. In theory, the perturbations must only balance any noise in the oscillator; practically, the controller must also overcome noise in the measurement system and any errors or deficiencies in the control implementation. In an experimental demonstration of chaos control, it is important to show that the control perturbations are truly minimal if one is to believe that the controlled orbit is a natural state of the uncontrolled system. Often, a control parameter can be tuned and the relative strength of the control perturbation can be monitored. If so, the presence of a well-defined local minimum in the observed control power as the parameter is tuned is a strong indication of efficient chaos control.

## 2. Chaotic Tuned Collector Transistor Oscillator

The schematic of our chaotic RF transistor oscillator is shown in Figure 1. The circuit is based on a standard tuned collector transistor oscillator that has long been used for RF applications [Hester, 1968]. The standard oscillator is augmented with a nonlinear folding circuit, which is implemented using a diode switch to drive a transistor-based integrator. The folding circuit emulates the nonlinear element of Carroll's low-frequency piecewise-linear circuit [Carroll, 1995], which exhibits a chaotic attractor similar to Rossler's simply folded band [Rossler, 1976]. The entire circuit was constructed on a breadboard using discrete, commonly available components. The RF coil was custom wound using 22-guage magnet wire on an 8-mm diameter air core. The primary winding (*i.e.*, the tank inductor) consists of five turns, while the secondary winding consists of two turns. Not shown in the schematic is a tertiary winding, which consists of two turns and provides an ac-coupled output signal to monitor the tank oscillations using an oscilloscope. The spacing of the turns in the primary winding is adjusted so that the uncontrolled circuit oscillates near 45 MHz.

For the oscillator, the transistor bias voltage, which is adjusted with the potentiometer, sets the feedback gain in the oscillator. For moderate gain, the circuit exhibits simple periodic oscillations. As the gain is increased, the oscillator shows a period-doubling sequence leading to chaos. A typical chaotic waveform generated by this circuit is shown in Figure 2. This waveform was captured using a digital sampling oscilloscope connected to the coil's tertiary winding, from which the tank voltage is deduced. A projection of the attractor is shown in Figure 3, which is generated from the captured waveform using a time-delay embedding ($\Delta t$ = 5 ns). Similar to Carroll's circuit, this autonomous oscillator exhibits chaotic dynamics characterized by a simply folded band attractor.

## 3. Limiter Control

To control chaos in this RF oscillator, we implement limiter control by attaching a controller circuit to the oscillator as shown in Figure 4. The efficacy of limiter control has been established previously, including application of a diode limiter to control chaos in an electronic circuit [Corron *et al.*, 2000]. Compared to other closed-loop chaos control techniques, limiter control is well suited for RF systems due to the minimal feedback latency that can be attained with a diode limiter.

In Figure 4, the emitter-base junction of a PNP transistor is used as a diode limiter restricting the dynamic range of the oscillator's tank voltage, $v_C$. The voltage $V_{LIM}$ sets the limiter level and is used to tune the controller. In operation, the limiter is set such that the emitter-base junction is usually reverse biased and, for most of the cycle, the oscillator dynamics are unaffected by the limiter. However, at waveform minima when $v_C$ falls below $V_{LIM} - V_D$ (where $V_D \sim 0.7$ V), the junction becomes forward biased and a control current is injected into the LC tank. Since $v_C$ is below this threshold only for the minimum troughs of the oscillator waveform, the control perturbation consists of a sequence of current pulses. The amplitude and duration of the pulses depends on the vigor with which the system state tries to penetrate the limiter. Equivalently, a discrete diode may be used for the limiter; however, a transistor amplifies the current pulses and provides a convenient voltage output $v_T$ to monitor the relative control signal without loading the oscillator.

As $V_{LIM}$ is adjusted, the controlled oscillator responds with various chaotic and periodic states; in fact, the sequence of observed states is consistent with the bifurcation diagrams obtained by Glass and Zeng in limiting one-dimensional maps [Glass and Zeng, 1994]. In general, the observed states are not natural orbits of the uncontrolled oscillator, since perturbations due to the limiter significantly distort the oscillator dynamics. However, at $V_{LIM} = 3.14$ V, the control perturbations diminish significantly and the system exhibits the period-1 orbit shown in Figure 5. The frequency of this periodic waveform is 43.9 MHz. In Figure 6, the peak magnitude of the control current pulses—amplified at $v_T$ and averaged over a 2-μs window—is plotted as a function of limiter voltage. The period-1 orbit shown in Figure 5 corresponds to a well-defined local minimum in this plot, thereby providing strong evidence that this orbit corresponds to the period-1 UPO for the uncontrolled oscillator.

Practically, a small deviation from the true UPO certainly exists due to non-ideal characteristics of the diode limiter. For ideal limiter control, we expect that the period-1 UPO can be stabilized with diode currents approaching the noise level of the system. However, for the voltage range of our LC oscillator, simple junction diodes do not turn on sharply as the applied voltage exceeds the threshold, resulting in soft limiting that requires higher control currents than would be necessary with an ideal diode [Wagner and Stoop, 2001]. In Figure 6, the effect of a non-ideal limiter is seen in the period-1 minimum that exceeds the noise floor of the system. A truer representation of the period-1 orbit would require a better limiter, *i.e.*, a diode that exhibits a sharper transition in its current-voltage characteristic.

## 4. Conclusions

In this letter, we show experimental results demonstrating UPO control of chaos at 43.9 MHz using limiter control. This is the fastest chaos control demonstration reported in the literature to date, further demonstrating that controlling chaos in RF systems is technically achievable. Consequently, we believe that chaos control in

general—and limiter control in particular—may enable practical implementation of engineering applications of chaos, including communications using symbolic dynamics.

**References**

T. L. Carroll [1995] “A simple circuit for demonstrating regular and synchronized chaos,” *Am. J. Phys.* **63**(4), 377-379.

N. J. Corron, S. D. Pethel, and B. A. Hopper [2000] “Controlling chaos with simple limiters,” *Phys. Rev. Lett.* **84**(17), 3835-3838.

A. Garfinkel, M. L. Spano, W. L. Ditto, and J. N. Weiss [1992] “Controlling cardiac chaos,” *Science* **257**, 1230-1235.

L. Glass and W. Zeng [1994] “Bifurcations in flat-topped maps and the control of cardiac chaos,” *Int. J. Bifurcation Chaos* **4**(4), 1061-1067.

S. Hayes, C. Grebogi, and E. Ott [1993] “Communicating with chaos,” *Phys. Rev. Lett.* **70**(20), 3031-3034.

S. Hayes, C. Grebogi, E. Ott, and A. Mark [1994] “Experimental control of chaos for communication,” *Phys. Rev. Lett.* **73**(13), 1781-1784.

D. L. Hester [1968] “The nonlinear theory of a class of transistor oscillators,” *IEEE Trans. Circuit Theory* **15**(2), 111-118.

K. Myneni, T. A. Barr, N. J. Corron, and S. D. Pethel [1999] “New method for the control of fast chaotic oscillations,” *Phys. Rev. Lett.* **83**(11), 2175-2178.

E. Ott, C. Grebogi, and J. A. Yorke [1990] “Controlling chaos,” *Phys. Rev. Lett.* **64**(11), 1196-1199.

K. Pyragas [1992] “Continuous control of chaos by self-controlling feedback,” *Phys. Lett. A* **170**(6), 421-428.

O. E. Rossler [1976] “An equation for continuous chaos,” *Phys. Lett. A* **57**(5), 397-398.

J. E. S. Socolar, D. W. Sukow, and D. J. Gauthier [1994] “Stabilizing unstable periodic orbits in fast dynamical systems,” *Phys. Rev. E* **50**(4), 3245-3248.

C. Wagner and R. Stoop [2001] “Optimized chaos control with simple limiters,” *Phys. Rev. E* **63**(1), 017201.

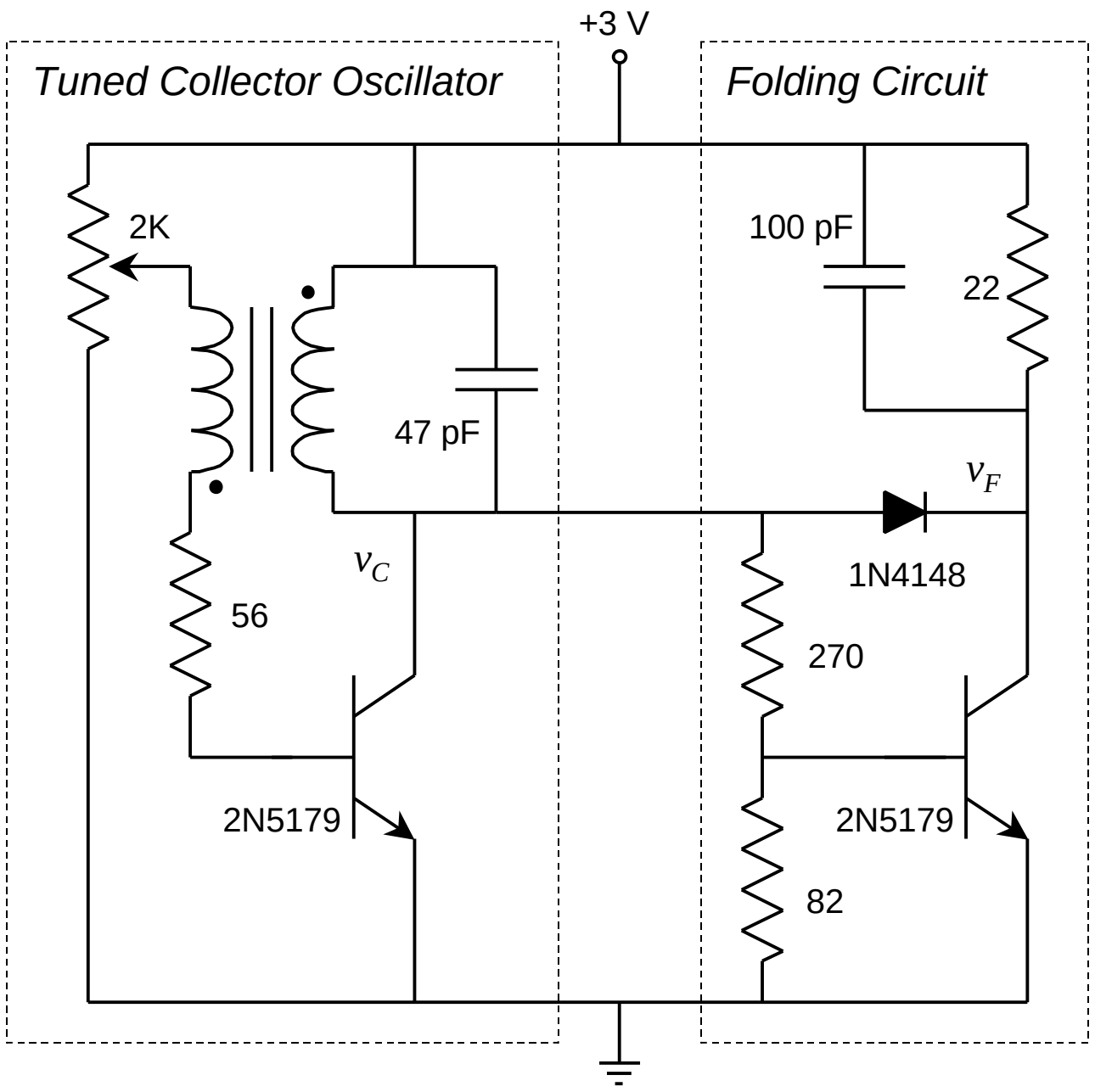

Figure 1. Chaotic tuned collector transistor oscillator. This circuit uses a standard RF transistor oscillator design augmented with a nonlinear folding circuit.

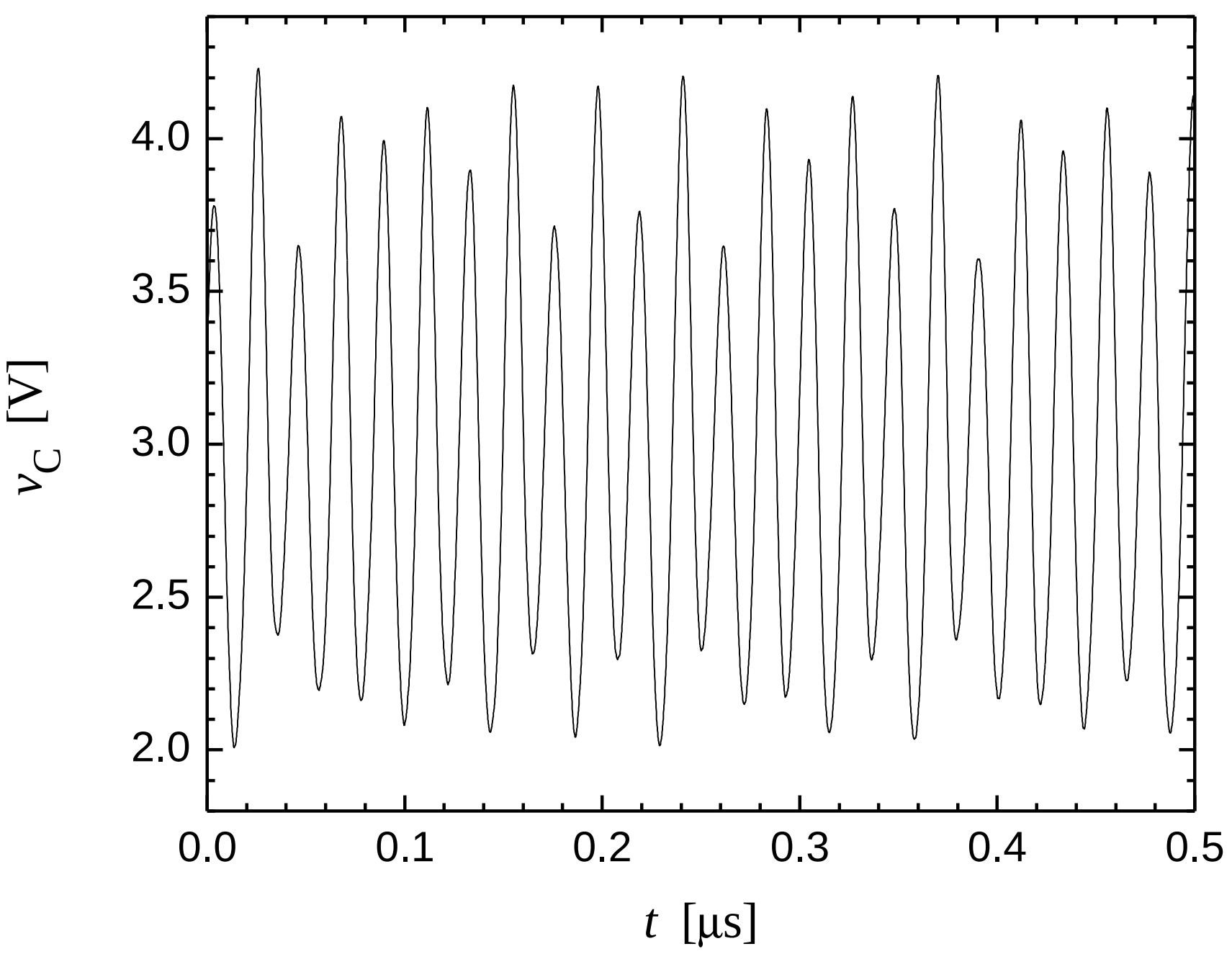

Figure 2. Typical waveform generated by the chaotic tuned collector transistor oscillator without control.

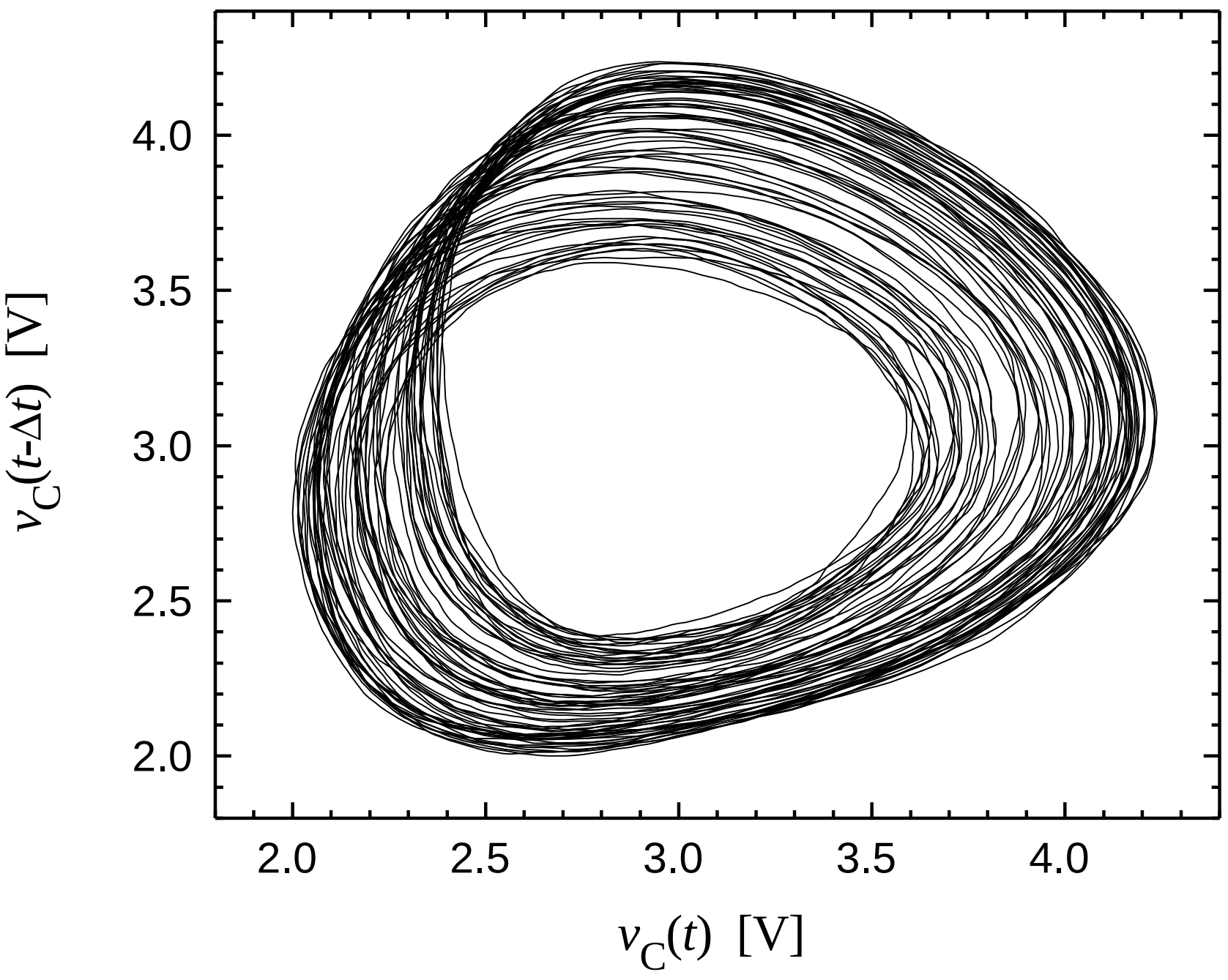

Figure 3. Attractor generated by the uncontrolled tuned collector oscillator using time-delay embedding ($\Delta t$ = 5 ns).

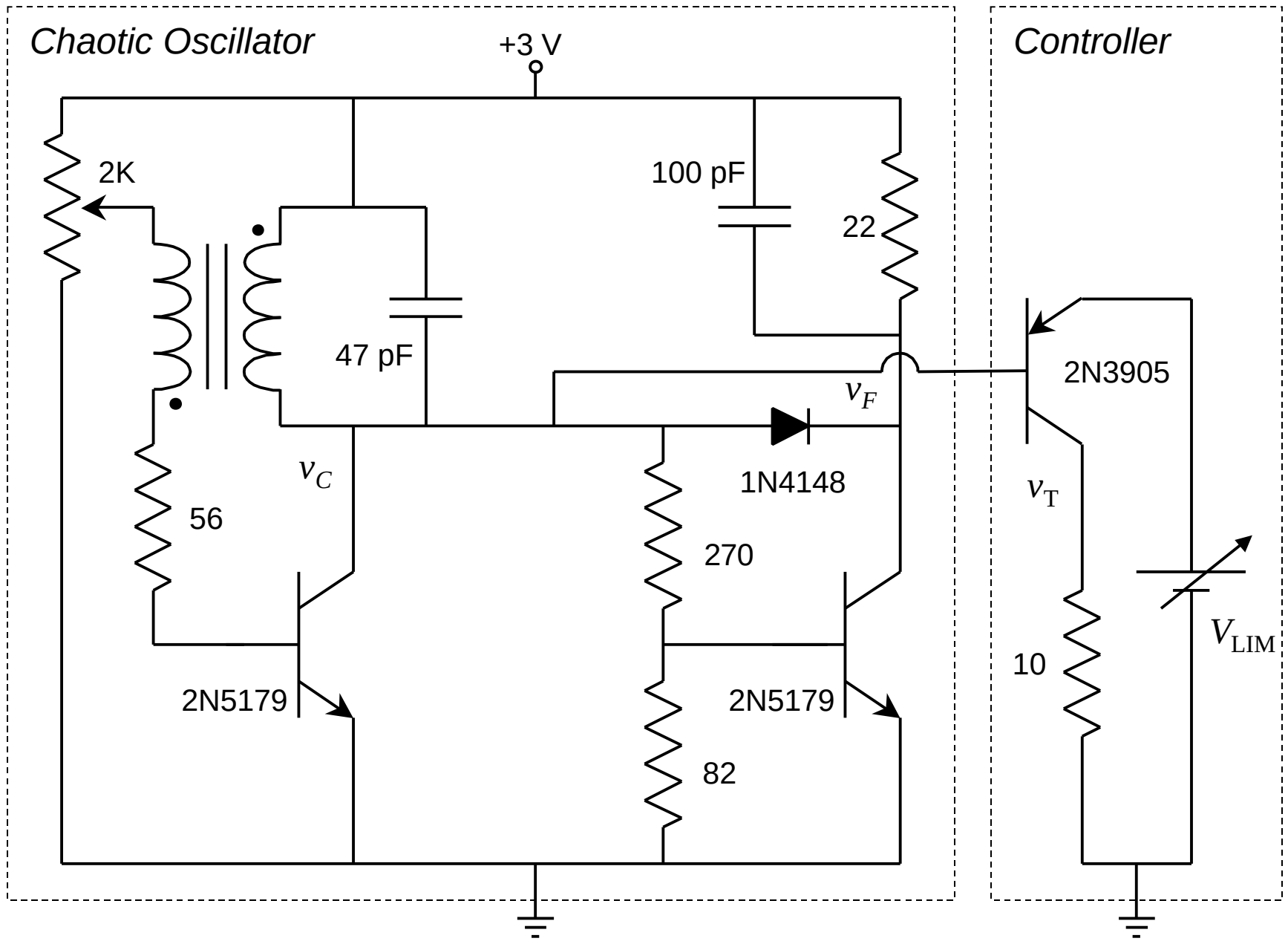

Figure 4. Diode limiter controller implemented using the emitter-base junction of a PNP transistor (2N3905) and applied to the chaotic tuned collector transistor oscillator. The voltage $V_{LIM}$ sets the limiter level applied to the tank voltage $v_C$.

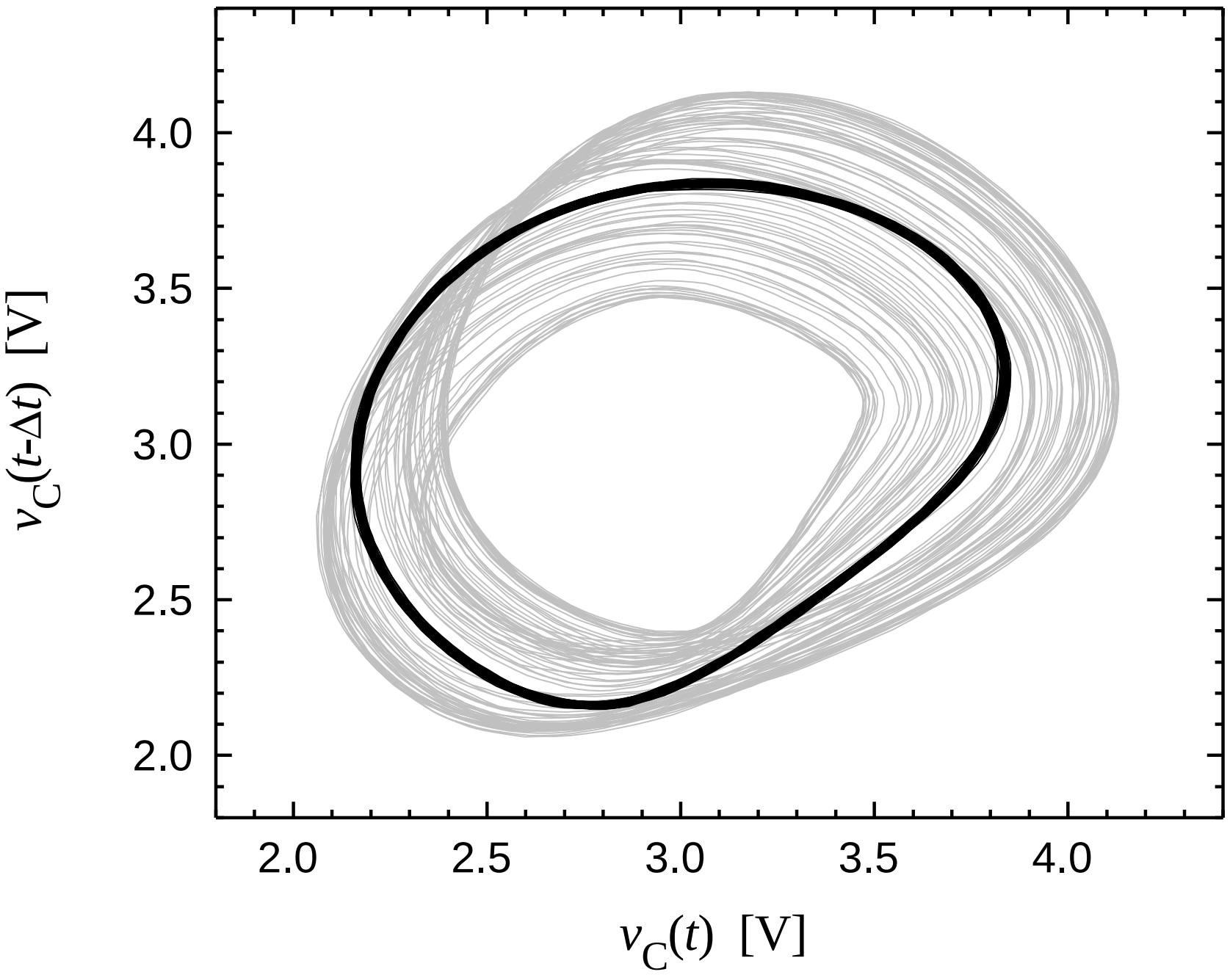

Figure 5. Stabilized period-1 UPO for the tuned collector oscillator compared to attractor of the uncontrolled oscillator. The frequency of the periodic waveform is 43.9 MHz ($\Delta t$ = 5 ns).

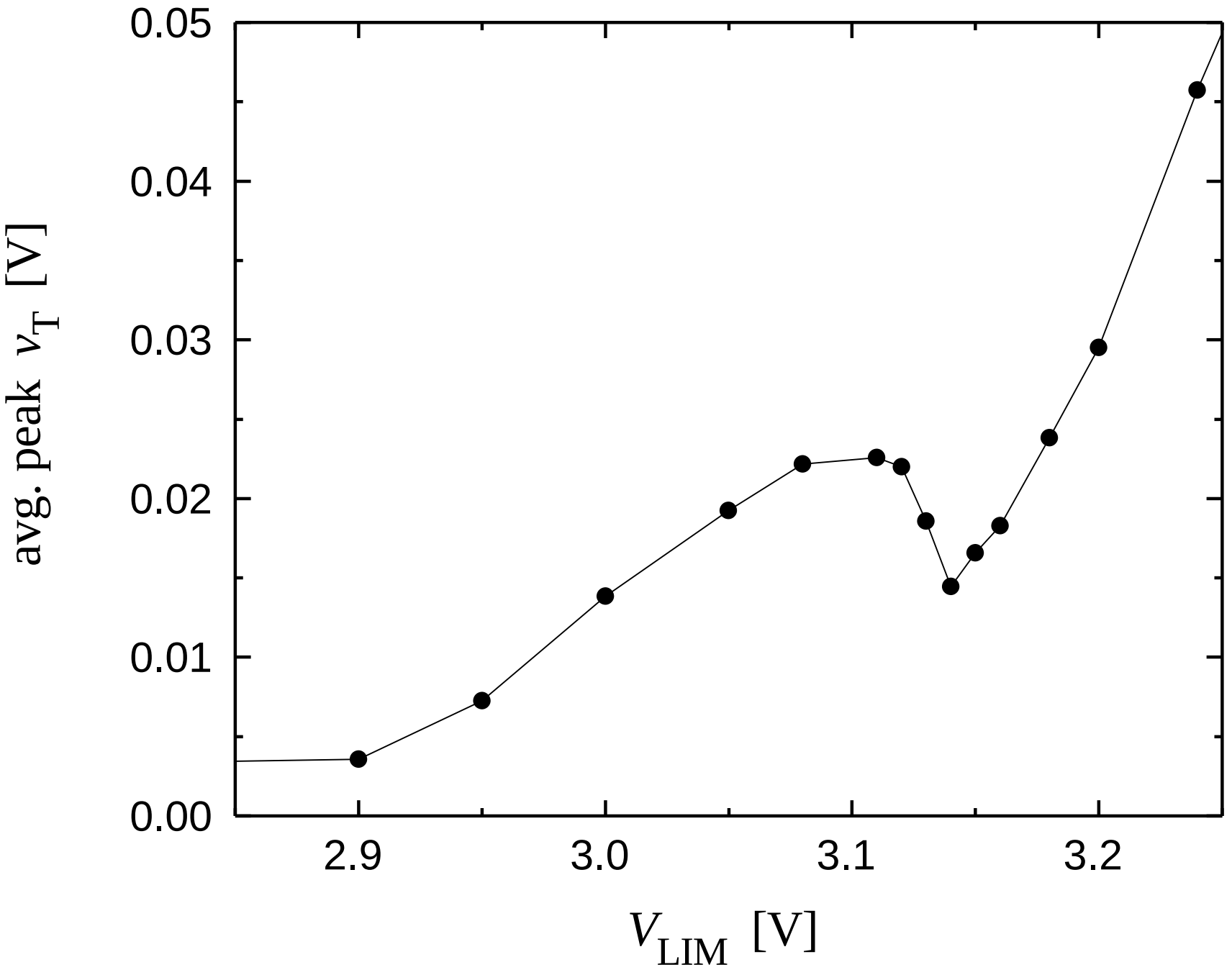

Figure 6. A measure of the peak control signal, averaged over a 2-μs window, to stabilize periodic orbits in the chaotic tuned collector oscillator. The local minimum at $V_{LIM}$ = 3.14 V corresponds to the stabilized period-1 UPO.